\documentclass[12pt]{article}
\usepackage{graphicx,epsfig}
\newcommand{\be}{\begin{equation}}
\newcommand{\ee}{\end{equation}}
\newcommand{\ba}{\begin{array}}
\newcommand{\ea}{\end{array}}

\renewcommand{\thefootnote}{\fnsymbol{footnote}}

\def\bea{\begin{eqnarray}}
\def\eea{\end{eqnarray}}

\begin{document}
\renewcommand{\thefootnote}{\fnsymbol{footnote}}
\rightline{IPPP/02/71} \rightline{DCPT/02/141} \rightline{KEK/TH/871} 
\vspace{0.3cm}
\begin{center}
{\Large Exclusive $J/\psi$ Productions at $e^+\ e^-$ Colliders}\vspace{0.5cm}\\
{\normalsize Kaoru Hagiwara$^1$, Emi Kou$^2$,  
and Cong-Feng Qiao$^{3,4}$} \\
{\small $^1$Theory Group, KEK, Tsukuba, Ibaraki 305-0801, Japan\\[2mm]
$^2$Institute of Particle Physics Phenomenology, \\
University of Durham, Durham, DH1, 3LE, UK\\[2mm]
$^3$Department of Physics, the Graduate School of the Chinese 
Academy of Sciences,\\
YuQuan Road 19A, 100039 Beijing, China\\[2mm]
$^4$The Abdus Salam International Centre for Theoretical 
Physics(ICTP),\\
Strada Costiera, 11-34014 Trieste, Italy}
\end{center}
\vspace{1cm}

\begin{abstract}

\noindent
Exclusive quarkonium pair production in electron-positron
collisions is studied in non-relativistic QCD. The obtained  cross section
for $J/\psi$ $\eta_c$ production in the leading order is confronted against the recent 
measurements by the Belle Collaboration at KEKB. It is shown that 
a large renormalization $K$-factor is necessary to explain the experimental data. 
We point out that the $J^{PC}=0^{-+}$ nature of the hadronic systems that are assigned to 
be $\eta_c$ should be tested by the triple angular distributions in terms of the 
scattering angle, and, polar and azimuthal angles of $J/\psi$ into leptons.  
We further study $J/\psi\  J/\psi$ and $\Upsilon\ \Upsilon$ productions at LEP energies. 
Although the axial-vector couplings of the $Z$-boson to charm and bottom quarks allow 
production of such pairs when one of them is polarised transversally and the other 
longitudinally, we find that the integrated luminosity at $Z$ pole accumulated by LEP 
is not large enough to observe the exclusive pair production of  quarkonium.   
\end{abstract}

\newpage
\section{Introduction}
Quarkonium production and decays have long been considered as an ideal 
means to investigate the bound-state formation in QCD. 
Due to its approximately non-relativistic nature, the description 
of the heavy quark and anti-quark system is one of the 
simplest applications of QCD. For example, the calculation of quarkonium 
leptonic decays render experimental results with a high precision, 
which may play a crucial role in investigating 
various phenomena such as measuring the parton distribution, 
detecting the Quark-Gluon-Plasma signal and even new physics. 

While the quarkonium physics has been studied for more than twenty years, 
the recent interest in the field has been focused on the colour-octet scenario 
\cite{fleming} which 
was triggered by the high-$p_T$ $J/\psi$ surplus production 
discovered by the CDF collaboration at the Tevatron in 1992 \cite{cdf1,cdf2,cdf3}.  
It was proposed based on a novel effective theory, the non-relativistic QCD(NRQCD) \cite{nrqcd}. 
Having achieved the first-step of explaining the CDF data, 
the colour-octet mechanism(COM) had a strong impact into the quarkonium physics and 
various efforts have been made to confirm this mechanism. Although the theoretical 
framework seems to show qualitative agreements with experimental data, 
there are certain difficulties in the quantitative estimate of the colour-octet contribution \cite{qcf1},  
in particular, in HERA physics  \cite{rothstein}. 
It was in such circumstances that the B factory experiments reported their first 
result on the prompt charmonium production from $e^+e^-$ collider at $\sqrt{s}=10.6$ GeV
\cite{BABAR1,BELLE1}. As far as hadronic uncertainty is concerned, the B factories 
would provide clearer information of the quarkonium production. 

The first result for the inclusive 
$e^+ + e^-\rightarrow J/\psi + X$ process from Belle (with $32.7fb^{-1}$ data set) 
indicated a discrepancy from the theoretical prediction \cite{BELLE1}. 
The $e^+ + e^-\rightarrow J/\psi + c\bar{c}$ process seemed to dominate the threshold region 
($z\to1$) of the energy spectra of the differential cross section 
comparing to the COM process $e^+ + e^-\rightarrow J/\psi + g$  and colour-singlet process 
$e^+ + e^-\rightarrow J/\psi + gg$, contrary to the theoretical expectation in \cite{tadegrand, ChoLeibo, Ko, fyuan, vvkiselev}. 
In their second report (based on $41.8fb^{-1}$ data set) \cite{BELLE2}, 
the direct measurement of the $e^+ + e^-\rightarrow J/\psi + c\bar{c}$ process is presented 
simultaneously and it is found that experimental result is 
about 10 times larger than the theoretical prediction for this process. 
In \cite{BELLE2}, the total cross section of the exclusive $e^+ + e^-\rightarrow J/\psi+ \eta_c$ 
process is found to be: 
\begin{equation}
\sigma (e^+ + e^-\rightarrow J/\psi+ \eta_c)\times \mathcal{B}(\eta_c\to \ge 4\mbox{charged})
=(0.033^{+0.007}_{-0.006}\pm 0.009) \mbox{pb} \label{eq:1}
\end{equation}
Note that the observed number of events for $J/\psi\ \eta_c$ is $(67^{+13}_{-12})$. 
Motivated by this measurement, we investigate $J/\psi\ \eta_c$ production at $e^+e^-$ colliders.

The reminder of the paper is organized as follows. In Sec. 2, we will give 
all the formulae used in our analysis. Motivated by the observed large cross section 
of the $J/\psi\ \eta_c$ production by Belle, 
we further consider the heavy quarkonium pair production at LEP energies. For this purpose, we 
include the formulae for the $Z$ intermediated processes.  
In Sec. 3, our numerical results are presented. The double $J/\psi$ production through 
$\gamma^* \gamma^*$ intermediated states which has been proposed as an explanation of  the large cross section of 
the $J/\psi\ \eta_c$ process ~\cite{bodwin2} is also examined.  
Finally we give our summary and conclusions in Sec. 4. 

\section{Formulae}
In this section, we give the formulae which we use in the following sections: 
the $e^+e^-$ annihilation into a pair of  
$1S$ charmonium- and bottomonium-sates. 
We use the standard method in our calculation: 
we start from the double $c\bar{c}$ (or $b\bar{b}$) production amplitudes,  
and project out the heavy quark and anti-quark pairs into the 
S-wave states in the colour-singlet (see Fig. 1). 

The spin projection operator for the quarkonium production is given by
\begin{equation}
\label{proj}
{\cal P}_{S,S_z}(P; q) = \sum_{s_1, s_2} v(\frac{P}{2} -q; s_2) \bar{u}
(\frac{P}{2} + q; s_1) <\frac{1}{2}, s_1; \frac{1}{2}, s_2|S, S_z>\ ,
\end{equation}
where $P$ and $S$, $S_z$ are respectively the quarkonium four-momentum, 
its spin and the $z$ component of the spin; $q$ is the relative 
momentum of the heavy quarks; and $s_1$, $s_2$ 
represent their spins. In  the  non-relativistic limit, 
the covariant forms of the projection operators are very simple: 
\begin{eqnarray}
{\cal P}_{0,0}(P; 0) = \frac{1}{2\sqrt{2}}\gamma_5 (\not\!{P} + M) \label{proj0} \\
{\cal P}_{1,S_z}(P; 0) = \frac{1}{2\sqrt{2}}\not\!{\epsilon^*}(P, S_z) (\not\!{P} + M)
\label{proj1}
\end{eqnarray}
respectively, for the  pseudoscalar and the vector quarkonium. 
Here $\epsilon^{\mu}(P, S_z)$ denotes the polarization vector of 
the spin-1 quarkonium state, and $M=2m$ is the mass of the quarkonium.
Projectors (\ref{proj0}) and (\ref{proj1})
map a $Q\bar{Q}$ pair into the S-wave states. 
In the rest frame of the vector meson, the polarization vector is given by
\begin{equation}
\label{polv}
\epsilon^{\mu}_0 \equiv \epsilon^{\mu}(P;S_z=0)= (0, 0, 0, 1)\ ,\ 
\epsilon^{\mu}_{\pm} \equiv \epsilon^{\mu}(P;S_z=\pm1)= (0, \mp, -i, 0)/\sqrt{2}\ ,
\end{equation}
for the longitudinal and transverse polarizations, respectively. 
We need to boost the polarisation vector from rest frame to the 
laboratory system along the spin quantization axis so that $S_z$ denotes the helicity 
($\lambda$) in the laboratory frame. In the frame where the electron beam is along the $z$-axis 
and the quarkonium scattering angle is $\theta$, the polarization vectors read  
\begin{equation}
\label{polv1}
\epsilon^{\mu}_0 = \gamma (\beta, \sin\theta, 0, \cos\theta)\ ,\ 
\epsilon^{\mu}_{\pm} = (0, \mp\cos\theta, -i, \pm\sin\theta)/\sqrt{2}\ .
\end{equation}
respectively, where 
\begin{equation}
\beta =  \sqrt{1 - \frac{4 M ^2}{s}},\ \ \ \ \ 
\gamma = \frac{1}{\sqrt{1 - \beta^2}}.
\end{equation}

\begin{figure}
\begin{center}
\includegraphics[width=15cm]{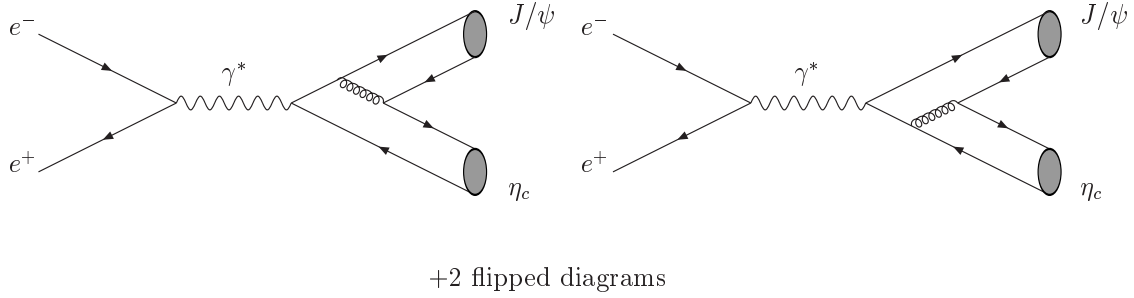}
\end{center}
\vspace*{-0.5cm}
\caption{Feynman diagrams for the $J/\psi\ \eta_c$ production from $e^+ + e^-$ 
annihilation.}
\end{figure}

Now let us show our analytical formulae for the cross sections. 
The helicity amplitudes for the process 
$ e^+ + e^- \rightarrow J/\psi + \eta_c(\Upsilon + \eta_b)$ is given by 
\begin{eqnarray}
M^{\sigma}_{\lambda=\pm} &=&-\frac{64g^{2}e^{2}R_Q(0)^2}{3\pi s^{3/2}}
\left[\frac{e_{Q}}{s}-
\frac{v_{Q}(v_{e}-\sigma a_{e})}{s-m_{Z}^{2}+im_{Z}\Gamma_{Z}}\right]
\epsilon^{\alpha\beta\mu\nu}\epsilon^{*}_{\lambda\alpha}P_{V_Q\beta}P_{\eta_Q\mu}j^{\sigma}_{\nu}  \label{note8} \\
M^{\sigma}_{\lambda=0} &=& 0 
\end{eqnarray}
where  $\sigma$ denotes the electron helicity in units of $1/2$, $\lambda$ the 
$J/\psi\ (\Upsilon)$ helicity, and $Q$ is $c$ or $b$. The initial current is 
given as $j^{\nu}_{\sigma}=(0,-i,\sigma ,0)$  
and $P_{V_Q}^{\mu}$ and $P_{\eta_Q}^{\mu}$ 
are the four momentum of vector and pseudoscalar mesons, respectively. 
The vector and axial-vector couplings of the $Z$-boson are 
\begin{equation}
\label{couplings}
v_e = -\frac{1 - 4 \sin^2\theta_w}{4 \sin\theta_w \cos\theta_w},\
a_e = -\frac{1}{4 \sin\theta_w \cos\theta_w},\
v_Q = \frac{I^{QL}_3 - 2 e_Q \sin^2\theta_w}
{2 \sin\theta_w \cos\theta_w}\ ,\
a_Q = \frac{I^{QL}_3}{2 \sin\theta_w \cos\theta_w}\ , 
\end{equation}
where the $I^{QL}_3$ is the third component of the weak isospin of the 
left-handed quark doublet; $e_Q$ is the $Q$ charge in units of proton charge. 
The absence of the $\lambda=0$ amplitudes, the scattering angle dependence of the 
$\lambda=\pm $ amplitudes as well as their relative sign tell us the spin parity 
($J^P=0^-$) of the hadronic system assigned as $\eta_c$. 

The above predictions can be tested by experiments through the decay angular distributions of 
$V_Q \to l^+l^-$, ($V_c=J/\psi, \ V_b=\Upsilon$). 
Using a definition 
$D_{\lambda}^{\sigma^{\prime}}\equiv M(V_{\lambda}\to l_{\sigma^{\prime}}\bar{l}_{\sigma^{\prime}})$
where $\sigma^{\prime}$ is the $l^-$ helicity in the $m_l=0$ 
limit ($l=e,\ \mu$), we obtain 
\begin{equation}
D_{\pm}^{\sigma^{\prime}}=\sqrt{\frac{3B}{16\pi}}(\sigma^{\prime}\pm\cos\theta^*)\frac{1}{\sqrt{2}}
e^{\mp i\sigma^{\prime}\phi^*}\label{note11}
\end{equation}
for $\lambda=\pm$ and 
\begin{equation}
D_{0}^{\sigma^{\prime}}=\sqrt{\frac{3B}{16\pi}}\sin\theta^*\label{note12}
\end{equation}
for $\lambda=0$. 
The normalization for the decay amplitudes Eqs. (\ref{note11}) and (\ref{note12}) is 
\begin{equation}
\sum_{\sigma^{\prime}}\int d\cos\theta^*d\phi^*|D^{\sigma^{\prime}}_{\lambda}|^2=B
\equiv B(V_Q \to l^+l^-) \label{eq:note2}
\end{equation}
Here $\theta^*$ and $\phi^*$ are the polar and azimuthal angles of $l^-$ in the $V_Q$ 
rest frame. The $\theta^*$ is measured from the $V_Q$ momentum direction in the $e^+e^-$ 
collision rest frame, and $\phi^*$ is measured from the scattering plane 
($\phi^*=\pi/2$ is along the $\vec{k}\times\vec{q}$ direction where $\vec{k}$ and 
$\vec{q}$ are, respectively the electron and $V_Q$ three momenta in the $e^+e^-$ 
collision c.m. frame). 
The triple angular distributions are then obtained as 
\begin{eqnarray}
\frac{d\sigma}{d\cos\theta d\cos\theta^*d\phi^*}&=&\frac{1}{2s}\frac{1}{4}\sum_{\sigma\sigma^{\prime}}
|\sum_{\lambda=\pm}M_{\sigma}^{\lambda}D_{\lambda}^{\sigma^{\prime}}|^2\frac{\beta}{16\pi} 
\nonumber \\
&=&\left(\frac{64}{3}\right)^2\frac{\pi \alpha^2 \alpha_s^2}{s^3}
\left[\frac{e_Q^2}{s}
-\frac{2 e_Q v_Q v_e (s - m_Z^2)-s v_Q^2 (a_e^2 + v_e^2)}{(s - m_Z^2)^2 + (m_Z \Gamma_Z)^2}\right]
|R_Q(0)|^4 \beta^3 \nonumber \\
&\ &\hspace{1cm}\times\frac{3B}{16\pi}
\left[(1+\cos^2\theta)(1+\cos^2\theta^*)-\sin^2\theta\sin^2\theta^*\cos 2\phi^*\right] 
\end{eqnarray}
where the initial beams are unpolarized and final lepton polarizations are unobserved. 
The $\cos\theta^*$ dependence tells us that only the transversally polarized $V_Q$ 
are produced and the $\phi^*$ dependence tells us the relative phase of the $\lambda=+$ 
and $\lambda=-$ amplitudes. 
These predictions of the $V_Q\ \eta_Q$ production processes should be tested experimentally. 
After integrating out the $V_Q$ decay angles and the scattering angle, we find the total 
cross section 
\begin{equation}
\frac{d\sigma}{d \cos\theta}=\left(\frac{64}{3}\right)^2
\frac{\pi \alpha^2 \alpha_s^2}{s^3} 
\left[
\frac{e_Q^2}{s}-\frac{2 e_Q v_Q v_e (s - m_Z^2)-s v_Q^2 (a_e^2 + v_e^2)}{(s - m_Z^2)^2 + (m_Z \Gamma_Z)^2}\right]
|R_Q(0)|^4 \beta^3 (1 + \cos^2\theta)
\label{eq:9} 
\end{equation}

We note in passing that a pair of $V_Q$ (e.g. $J/\psi+J/\psi$ or $\Upsilon+\Upsilon$) 
cannot be produced from a single $\gamma^*$ state because of charge conjugation invariance. 
Such pairs can be produced via $2\gamma^*$ intermediate states or in the $Z$-boson 
decays through its axial-vector couplings to the heavy quarks. 
The former process has been studied in Ref. \cite{bodwin2}. For the latter process, we find 
the helicity amplitudes 
\begin{equation}
M^{\sigma}_{\lambda_1 \lambda_2}=
\frac{32g^2e^2a_QR_Q(0)^2M}{3\pi s^{3/2}}
\frac{(a_e-\sigma v_e)}{(s-m_Z^2+im_Z\Gamma_Z)}
\epsilon_{\alpha\beta\mu\nu}\epsilon^{*\alpha}_{\lambda_1}\epsilon^{*\beta}_{\lambda_2}
(P_{V_Q1}^{\mu}-P_{V_Q2}^{\mu})j^{\sigma \nu} \label{note19}
\end{equation}
where $\sigma$ is the electron helicity 
(in units of 1/2), $\lambda_1$ and $\lambda_2$ are the helicities of $V_Q$ with 
$\cos\theta > 0$ and $\cos\theta < 0$, respectively, 
$j_{\sigma}^{\mu}=(0,i\sigma , 1,0)$ is the initial $e^+e^-$ current, and 
$P_{V_Q1, 2}$ is the momentum of $V_Q$'s. 
From this result, 
one can easily find that the amplitudes for 
$(\lambda_1, \lambda_2)=(\pm, \pm), (0, 0), (\pm, \mp)$ are zero.  
The absence of the $\lambda_1=\lambda_2=\pm$ amplitudes is in accordance with the 
Yang's theorem (that forbids the transition between the spin 1 state and a pair of 
identical massless vector bosons), the $\lambda_1=\lambda_2=0$ amplitude is absent due to 
Bose symmetry ~\cite{Hikasa} and the $\lambda_1=-\lambda_2=\pm$ amplitudes vanish because of angular 
momentum mismatch. It is only a pair of longitudinally and transversally polarized 
$V_Q$'s that can be produced via the $Z$-boson exchange in $e^+e^-$ annihilation. 

The differential cross section after summing over the $V_Q$ helicities for 
unpolarized beams is 
\begin{equation}
\label{2jpsi}
\frac{d\sigma}
{d \cos\theta}\ =\ \left(\frac{32}{3}\right)^2\ 
\frac{\pi \alpha^2 \alpha_s^2}{s^{2}}
\frac{a_Q^2 (a_e^2 + v_e^2)}{(s - m_Z^2)^2 + (m_Z \Gamma_Z)^2}
\ |R_Q(0)|^4\ \beta^5\ (1 + \cos^2\theta)\ .
\end{equation}
From Eqs. (\ref{note8}) and (\ref{note19}), we find the $Z$-boson decay widths: 
\begin{eqnarray}
\Gamma (Z\rightarrow V_Q \eta_{Q}) &=&\left(\frac{64}{3}\right)^2
\frac{2\alpha\alpha_s^2v_Q^2|R_Q(0)|^4\beta^3}{3 m_Z^5}\\
\Gamma (Z\rightarrow V_Q V_{Q}) &=& \left(\frac{32}{3}\right)^2
\frac{2\alpha\alpha_s^2a_Q^2|R_Q(0)|^4\beta^5}{3 m_Z^5}
\end{eqnarray}

\section{Numerical Results}
The following input parameters are used in our numerical analysis in this section: 
\begin{eqnarray}
&\alpha=\frac{1}{137}, \ \ 
I_c=\frac{1}{2}, \ \ 
I_b=-\frac{1}{2}, \ \ 
M_Z=91.2\mbox{GeV}, \ \ 
\Gamma_{Z}=2.495\mbox{GeV}, \ \ 
\sin^{2}\theta_w=0.231, \ \ &\nonumber \\
&\Gamma_{J/\psi\to e^+e^-}=(5.26\pm 0.37)\times 10^{-6}\mbox{GeV}, \ \ 
\Gamma_{\Upsilon\to e^+e^-}=(1.32\pm 0.05)\times 10^{-6}\mbox{GeV}  & \label{eq:para} \\
&M_{J/\psi}=3.1\mbox{GeV}, \ \ 
M_{\Upsilon}=9.5\mbox{GeV} \ \ &\nonumber
\end{eqnarray}
Although our analysis is strictly in the leading order of perturbative QCD, we adapt the two-loop 
running coupling constant of $\overline{\mbox{MS}}$ scheme in order to define the ``K'' factor 
between the leading order prediction and the observed cross section. 
More specifically, we adopt 
\begin{equation}
\frac{\alpha_{s}}{4\pi}
\left(\mu\right)_{\overline{\mbox{\tiny MS}}}=
\frac{1}{\beta_0\ln (\mu^2/\Lambda^2_{\overline{\mbox{\tiny MS}}})}-
\frac{\beta_1\ln\ln(\mu^2/\Lambda^2_{\overline{\mbox{\tiny MS}}})}
{\beta_0^3\ln^2(\mu^2/\Lambda^2_{\overline{\mbox{\tiny MS}}})}
\end{equation}
where $\beta_0^{N_f}=(33-2N_f)/3, \beta_1^{N_f}=102-10N_f-8N_f/3$.  
$\alpha_{s}(M_{Z})_{\overline{\mbox{\tiny MS}}}=0.118$ corresponds to $\Lambda_{\overline{\mbox{\tiny MS}}}=0.226$ GeV 
for $N_f=5$. 
We set $\mu=\sqrt{s}/4$ as our leading order estimate, because the invariant mass of the 
exchanged gluons in the diagram of Fig. 1 is $\mu=\sqrt{s}/2$, and we account for the 
factor of two mismatch between the momentum scale and the $\overline{\mbox{MS}}$ scale \cite{BL++}. 
The value of the wave function at the origin can be extracted from the leptonic widths $\Gamma (V_{Q}\rightarrow l^{+}l^{-})$:  
\begin{eqnarray}
|R_c(0)|^2&=&\frac{9M^2_{J/\psi}}{16\alpha^2}\Gamma_{J/\psi\to e^+e^-} \label{eq:14}\\
|R_b(0)|^2&=&\frac{9M^2_{\Upsilon}}{4\alpha^2}\Gamma_{\Upsilon\to e^+e^-}\label{eq:15}
\end{eqnarray}
Using the experimental values \cite{PDG},   
we obtain: 
\begin{equation}
|R_c(0)|^2=(0.53\pm 0.04)\mbox{GeV}^3, \ \ \ \ \ |R_b(0)|^2=(5.0\pm 0.2)\mbox{GeV}^3. 
\label{eq:16}
\end{equation}

First of all, we show our numerical result for the total cross section of 
$ e^+ + e^-\rightarrow J/\psi + \eta_c$ 
which has already been measured by the Belle collaboration. Using the central values of 
Eq. (\ref{eq:16}) and $\Lambda_{\overline{\mbox{\tiny MS}}}=0.226$ GeV, Eq. (\ref{eq:9}) gives  
\begin{equation}
\sigma(e^+ + e^-\rightarrow J/\psi + \eta_c)=0.0023 \mbox{pb}. 
\end{equation}
If all the hadronic systems are indeed $\eta_{c}$ decay products and if we set 
\begin{equation}\mathcal{B}(\eta_c\to \ge 4\mbox{charged hadrons})=0.6\pm 0.1 \label{eq:chargem}\end{equation} 
we find:
\begin{equation}
K(\sqrt{s}=10.6\mbox{GeV})\equiv\frac{\sigma(e^+ + e^-\rightarrow J/\psi + \eta_c)_{\mbox{exp.}}}
{\sigma(e^+ + e^-\rightarrow J/\psi + \eta_c)_{\mbox{th.}}}=(24 \pm 12)  \label{eq:Kfacr1}
\end{equation}
where the error includes only experimental ones, Eq. (\ref{eq:1}) and that in our estimate Eq. (\ref{eq:chargem}). 
 
Let us examine if the large $K$-factor in Eq. (\ref{eq:Kfacr1}) can be explained. 
It is well known that the leptonic width formulae Eqs. (\ref{eq:14}) and (\ref{eq:15}) suffers from large NLO corrections. As a result, our estimate of the value of the wave function at the 
origin could include large error.  For instance, one of the potential model calculations  
give  \cite{Buchmuller, E-Q95}  
\begin{equation}
|R_c(0)|^2=0.810\mbox{GeV}^3, \ \ \ \ \ |R_b(0)|^2=6.477\mbox{GeV}^3. \label{eq:17}
\end{equation}
By replacing (\ref{eq:16}) by (\ref{eq:17}), we have  a factor of $(81/53)^{2}$. If we use $\Lambda_{\overline{\mbox{\tiny MS}}}=0.296$ GeV for 
$\alpha_{S}(M_{Z})=0.123$ and change the scale $\mu=\sqrt{s}/8$ instead of our standard 
choice of $\mu=\sqrt{s}/4$, 
we find  $\alpha_{S}(\mu=\sqrt{s}/8)=0.41$ instead of $\alpha_{S}(\mu=\sqrt{s}/4)=0.29$. 
This gives rise to another  
factor of $(35/26)^{2}$. The product of the two factors is about $5.8$ which is still significantly smaller than the 
the value Eq. (\ref{eq:1}) indicated from the experiment.  There should be further large contributions in the 
production amplitude and/or the hadronic system should contain significant amount of non-$\eta_{c}$ 
contributions.

\begin{figure}
\begin{center}
\includegraphics[width=17cm]{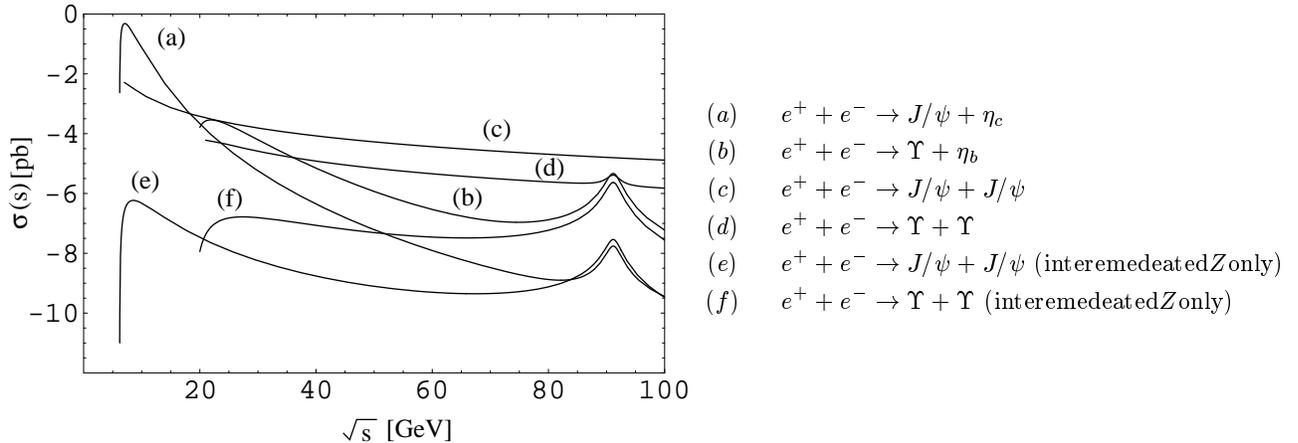}
\end{center}
\vspace{-0.5cm}
\caption{\small 
The $\sqrt{s}$ dependence of the total cross section for the exclusive 1S 
double-quarkonium production (in Log-scale) integrated over $|\cos\theta|\leq 0.9$ 
is shown.  The $K$-factor defined in Eq. (\ref{eq:22}) is multiplied for the 
$J/\psi\ \eta_c$ and $\Upsilon\ \eta_b$ production processes 
and the $J/\psi\ J/\psi$ and $\Upsilon\ \Upsilon$ processes from the intermediated $Z$ boson. } 
\label{fig:2}
\end{figure}

In Ref. \cite{bodwin2}, it has been shown that the $J/\psi$ pair production process via 
$\gamma^{*}\ \gamma^{*}$ intermediate state   
($e^+ + e^-\rightarrow \gamma^*\gamma^*\rightarrow J/\psi + J/\psi$) 
should be as large as (or larger than)  $e^+ + e^-\rightarrow  J/\psi + \eta_c$. 
Because $J/\psi\rightarrow \mbox{hadrons}$ can not be distinguished easily from   $\eta_{c}\rightarrow \mbox{hadrons}$,  
the Belle data Eq. (\ref{eq:1}) may contain contributions from $J/\psi$ pair production. The cross section  is estimated as: 
\begin{equation}
\sigma (e^{+}+e^{-}\rightarrow J/\psi+J/\psi )=0.0027 \mbox{pb} \label{eq:30-5}
\end{equation}
Only a small fraction of the observed cross section of Eq. (\ref{eq:1}) can 
come from the $J/\psi$ pair production process, because the normalization of the 
prediction Eq. (\ref{eq:30-5}) has little theoretical uncertainty. 
We note here that the $J/\psi$ pair production process has the following distribution properties: 
the pair of $J/\psi$ are  transversally (either $(+,-)$ or $(-,+)$) 
polarized and the differential cross section behaves as $(1+\cos^2\theta)/(1-\cos^2\theta)$  
while the $J/\psi\  \eta_c$ production behaves as $(1+\cos^2\theta)$ (see Eq. (\ref{eq:9})).  
The absence of interference between $\lambda=+$ and $\lambda=-$ amplitudes predicts that 
there is no azimuthal angle ($\phi^*$) dependence. The overall normalization of the $J/\psi$ 
pair production contribution should soon be found experimentally once the double leptonic 
decays of the $J/\psi$ pair are observed. 


Our numerical results are summarized in Fig. 2. 
In this plot, we multiplied the cross section for $J/\psi\  \eta_c$ production by 
the $K$-factor defined: 
\begin{equation}
K(\sqrt{s})=K(\sqrt{s_0})\times 
\left[\frac{\alpha_s(\sqrt{s/4})}{\alpha_s(\sqrt{s_0/4})}\right]^2
\label{eq:22}
\end{equation}
We use the same $K$ factor, with $K(10.6\ \mbox{GeV})=24$, for the single $\gamma^*$ and 
$Z^*$ exchange contributions to $J/\psi\ J/\psi$, $\Upsilon\ \eta_b$, and 
$\Upsilon\ \Upsilon$ production processes, even though the $K$ factor for the 
bottomonium pair production may be smaller than that for the charmonium pair production.    
We hoped that we could study $J/\psi\  \eta_c$ production  and related processes at 
$Z$ pole by using the LEP data. 
However considering the LEP integrated luminosity at $Z$ pole 
(about 1 fb$^{-1}$) and the branching ratio of the leptonic decay of $J/\psi$ (0.06), 
it is unfortunately impossible to observe this process at LEP. 
We find for the input parameters of Eq. (\ref{eq:para}) and $K(M_Z)=9.47$ from Eq. (\ref{eq:22}) 
the following branching fractions: 
\begin{eqnarray}
B(Z \rightarrow J/\psi + \eta_c) &=& 3.39\times 10^{-13} \\
B(Z \rightarrow J/\psi + J/\psi) &=& 5.73\times 10^{-13} \\
B(Z \rightarrow \Upsilon + \eta_b) &=&9.23 \times 10^{-11} \\
B(Z \rightarrow \Upsilon + \Upsilon) &=&4.61\times 10^{-11}. 
\end{eqnarray}
Even with the cut-off of $|\cos\theta|<0.9$, the $J/\psi$ pair production from two 
virtual photons dominate over all the other exclusive charmonium and bottomonium pair production 
processes at all energies except around the $B$ factory energies. This is essentially because 
the form factor of the exclusive heavy quarkonium production process drops sharply as $s^{-3/2}$ 
at high energies, as can be seen from Eq. (\ref{eq:9}): 
\begin{equation}
\frac{\sigma (e^+ + e^-\rightarrow J/\psi + \eta_c)}{\sigma (e^+ + e^-\rightarrow c\bar{c}) } 
\propto s^{-3}
\end{equation}
An extra suppression factor of $s^{-1}$ as compared to the high-energy behavior of the light 
meson pair production processes reflects the non-relativistic constraint that the two 
constituents should have the same velocity.

\section{Conclusions}
Exclusive $J/\psi\  \eta_c$ production in $e^+e^-$ collisions at $\sqrt{s}=10.6$ GeV 
is studied in view of the recent Belle observation \cite{BELLE2}.  
The observed total cross section turns out to be more than one 
order of magnitude larger than the naive leading order prediction 
of non-relativistic QCD, resulting in the huge renormalization 
factor of $K=24 \pm 12$.  We find that the $K$ factor of up to about 6 
can be obtained by taking account of the next-leading-order corrections 
to the $J/\psi$ leptonic width and by studying the scale dependence 
of the leading-order prediction.  On the other hand, the experimental 
data may contain contributions from non-$\eta_c$ origin hadronic events, 
such as hadrons from $J/\psi$ decays in the $J/\psi$-pair production via 
two virtual photon exchange \cite{bodwin2}, and hadrons from two gluon jets 
in the color-singlet $J/\psi + gg$ process \cite{tadegrand, ChoLeibo, Ko, fyuan, vvkiselev}.  
We propose to use the triple angular distribution of the $J/\psi$ production 
and $J/\psi$ decay into charged leptons to test the $J/\psi+\eta_c$ 
hypothesis.  A peculiar azimuthal angle dependence of the lepton 
distribution about the scattering plane is predicted.  

We have also studied $J/\psi +J/\psi$ production via $Z$-boson exchange 
and find that a pair of a transversally polarized $J/\psi$ and 
a longitudinally polarized $J/\psi$ can be produced in $Z$ boson 
decays via its axial-vector coupling to the charm quark.  
Unfortunately, the branching fraction of the $Z$ boson decays into 
$J/\psi +J/\psi$, $J/\psi +\eta_c$, $\Upsilon +\Upsilon$, $\Upsilon +\eta_b$  
are all too small to be observed in the LEP data, even with 
a possible large $K$ factor.  

Before closing this report, we note that pair production of 
$S$-wave and $P$-wave charmonium has been studied in Ref. \cite{bodwin1, chao}
and additionally $S$-wave + $D$-wave as well as $P$-wave + $P$-wave 
charmonium productions have been studied in Ref. \cite{bodwin1}.  We confirm 
their results of $J/\psi\ \chi_{cJ}$ (J=0,1,2) production cross sections.  
Although the perturbative calculation of the $J/\psi\ \eta_c$  
production cross section falls short of the observed one, 
it is still interesting to test whether the ratio among 
cross sections of all the above processes are consistent with 
the predictions of NRQCD.

\vspace{0.5cm}
\noindent
{\bf \large{Acknowledgements}} \\
We would like to thank Prafulla Kumar Behara, Jun-ich Kanzaki, Pasha Pakhlov and Bruce Yabsley 
for discussions about Belle data, and Zhi-Hai Lin for discussions about 
$J/\psi$ pair production via two virtual photons. 
The work of K. H. and E. K. were supported in part by the Monbu-Kagaku-sho 
Grant-in-Aid for  Scientific Research No.11202203. 
C.-F.Q. was supported in part by National 
Science Foundation of China with Contract No.19805015, and in part 
by the Core University Program between Chinese Academy of Science and Japan Society 
for Promotion of Science.   
E. K. and C.-F. Q. would like to thank the KEK theory group 
for their hospitality while this work was initiated.


\end{document}